\documentclass[apj]{emulateapj}
\usepackage{float,epsfig}
\usepackage{pstricks}
\newcommand{\etal}{\mbox{et~al.}}

\def\deg      {{\ifmmode^\circ\else$^\circ$\fi}} 

\slugcomment{ApJS 2007, COSMOS Special Issue}

 \shorttitle{The XMM-COSMOS Survey. I}
 \shortauthors{Hasinger \etal~}

\begin{document}
 
 \title{The XMM--Newton wide--field survey in the COSMOS field: I. Survey description}

 \author{ 
G. Hasinger\altaffilmark{1},
N. Cappelluti\altaffilmark{1},
H. Brunner\altaffilmark{1},
M. Brusa\altaffilmark{1},
A. Comastri\altaffilmark{2},
M. Elvis\altaffilmark{3},\\
A. Finoguenov\altaffilmark{1},
F. Fiore\altaffilmark{4},
A. Franceschini\altaffilmark{5},
R. Gilli\altaffilmark{2},
R. E. Griffiths\altaffilmark{6},
I. Lehmann\altaffilmark{1},\\
V. Mainieri\altaffilmark{1},
G. Matt\altaffilmark{7},
I. Matute\altaffilmark{1,16},
T. Miyaji\altaffilmark{6},
S. Molendi\altaffilmark{8},
S. Paltani\altaffilmark{9},
D. B. Sanders\altaffilmark{10},\\
N. Scoville\altaffilmark{11,12},
L. Tresse\altaffilmark{13},
C. M. Urry\altaffilmark{14},
P. Vettolani\altaffilmark{15},
G. Zamorani\altaffilmark{2}}

 
%
\altaffiltext{1}{Max Planck Institut f\"ur extraterrestrische Physik,  p.o. box 1312, D-85478 Garching, Germany}
\altaffiltext{2}{INAF-Osservatorio Astronomico di Bologna, via Ranzani 1, 40127 Bologna, Italy}

\altaffiltext{3}{Harvard-Smithsonian Center for Astrophysics, 60 Garden Street, Cambridge, MA 02138}
\altaffiltext{4}{INAF-Osservatorio Astronomico di Roma, Via Frascati 33, I-00044 Monteporzio Catone, Italy}
\altaffiltext{5}{Dipartimento di Astronomia, Universita di Padova, vicolo dell'Osservatorio 2, I-35122 Padua, Italy} 
%
%
\altaffiltext{6}{Department of Physics, Carnegie Mellon University, 5000 Forbes Avenue, Pittsburgh, PA 15213}
\altaffiltext{7}{Dipartimento di Fisica, Università degli Studi Roma Tre, via della Vasca Navale 84, 00146 Roma, Italy}
\altaffiltext{8}{INAF/IASF-CNR, Sezione di Milano, via Bassini 15, I-20133 Milan, Italy}

\altaffiltext{9}{INTEGRAL Science Data Centre, Chemin d'Ecogia 16, 1290 Versoix, Switzerland}
\altaffiltext{10}{Institute for Astronomy, 2680 Woodlawn Dr., University of Hawaii, Honolulu, Hawaii, 96822}
\altaffiltext{11}{California Institute of Technology, MC 105-24, 1200 East
California Boulevard, Pasadena, CA 91125}
\altaffiltext{12}{Visiting Astronomer, Univ. Hawaii, 2680 Woodlawn Dr., Honolulu, HI, 96822}
\altaffiltext{13}{Laboratoire d'Astrophysique de Marseille, BP 8, Traverse
du Siphon, 13376 Marseille Cedex 12, France}
\altaffiltext{14}{Department of Astronomy, Yale University, P.O. Box 208101, New Haven, CT 06520-8101}
\altaffiltext{15}{INAF-IRA, via Gobetti 101, 40129 Bologna, Italy}
\altaffiltext{16}{INAF-Osservatorio Astrofisico di Arcetri, Largo E. Fermi 5, 50125 Firenze, Italy}

  
 \begin{abstract}
We present the first set of XMM--Newton EPIC observations in the 2 square
degree COSMOS field. The strength of the COSMOS project is the unprecedented
combination of a large solid angle and sensitivity over the whole
multiwavelength spectrum.  The XMM--Newton observations are very efficient
in localizing and identifying active galactic nuclei (AGN) and clusters as
well as groups of galaxies.  One of the primary goals of the XMM--Newton
Cosmos survey is to study the co--evolution of active galactic nuclei as a
function of their environment in the Cosmic web. Here we present the log of
observations, images and a summary of first research highlights for the
first pass of 25 XMM--Newton pointings across the field.  In the existing
dataset we have detected 1416 new X--ray sources in the 0.5--2, 2--4.5 and
4.5--10 keV bands to an equivalent 0.5--2 keV flux limit of $7 \cdot
10^{-16}$ erg cm$^{-2}$ s$^{-1}$.  The number of sources is expected to grow
to almost 2000 in the final coverage of the survey.  From an X--ray color
color analysis we identify a population of heavily obscured, partially leaky
or reflecting absorbers, most of which are likely to be nearby,
Compton-thick AGN.

\end{abstract}
 
 
 

 
\section{Introduction}
 
COSMOS is a global multiwavelength collaboration built around an HST
Treasury Program providing deep images with the Advanced Camera for Surveys
(ACS) that cover an unprecedentedly large contiguous 2 deg$^2$ field
\citep{sco06}. One of the primary goals of COSMOS is to study the
co--evolution of galaxies and their central black holes out to high
redshifts, placing them in the context of the large scale structure in which
they reside and with high resolution morphological information. The
XMM--Newton observations are a crucial element of the COSMOS survey, because
of the superb efficiency of X--ray observations in localizing and
identifying active galactic nuclei (AGN) and distant clusters of galaxies.
 
\begin{figure*}
\epsscale{1.0} 
\plotone{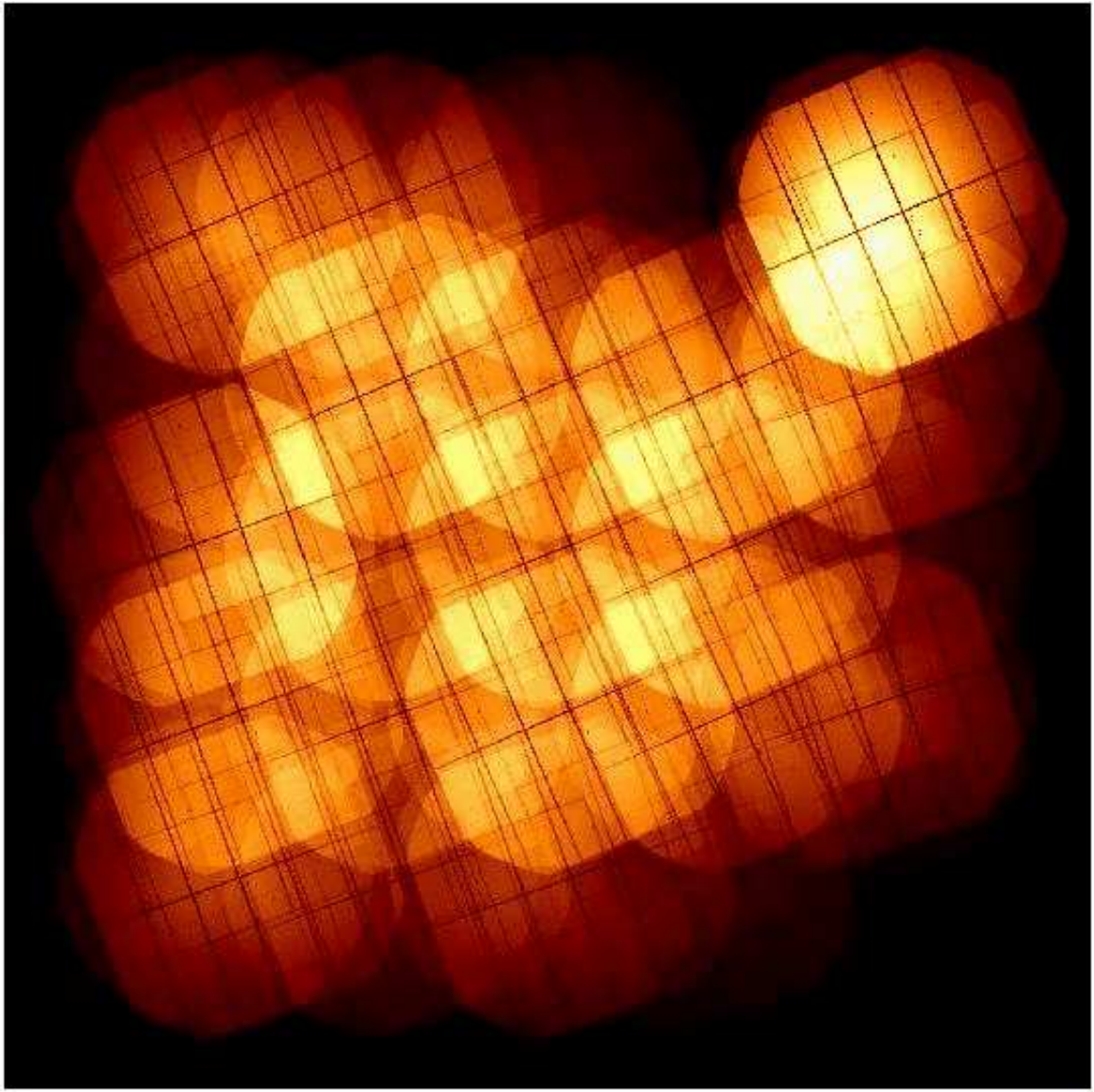}
\caption{Exposure map of the XMM--Newton raster scan in the COSMOS field.
The size of the field is about $1.4 \times 1.4$  deg$^2$, centered at
R.A.=$10^h00^m26.41s$; Dec.=$2d12\arcmin36\arcsec$. North is up and East to 
the left. The 25 individual pointings specified in Table \ref{tab:obs}
run along successive columns from North--East to South--West.} 
\label{fig:exp}
\end{figure*}

It has recently become clear that X--ray selected AGN are highly biased
tracers of the dark matter distribution, as shown by the significant degree
of angular auto--correlation and field to field variance in the source
counts \citep{cap01,cow02,yan03,alm03,eli04,cap05} as well as by the
presence of spikes in the AGN redshift distribution, mapping onto the
large--scale structure from optical data \citep{gil03,bar03}. The evolution
of the clustering with redshift and the comparison of the AGN clustering
signal with that of other galaxy populations gives important information on
the biasing of AGN and thus on the environment and type of dark matter halos
in which activity is triggered.  \citet{laf98} reported the tentative
detection of an increase of clustering of optically selected AGN with
redshift. The 2dF AGN survey found a factor of 2.7 increase of the
clustering amplitude from $z=0.5$ to $z=2.5$ \citep{cro05}, based on a
sample of more than 20000 objects.  The spatial clustering of X--ray
selected AGN has recently been measured in different fields
\citep{mul04b,gil05,yan06}, but the data so far do not allow to obtain
information on clustering evolution with redshift.

For a given survey area, the signal--to--noise ratio of the clustering
signal scales with the surface density (for 2D correlation) or the volume
density (for 3D correlation) of the objects in the sample.  AGN are very
rare objects, however, deep X--ray surveys are known to yield the largest
surface density of AGN in any waveband (e.g. 1000 deg$^{-2}$ at the faintest
limiting flux aimed for in the XMM--Newton COSMOS survey). Moreover, because
X--rays can both penetrate obscuring dust and are not sensitive to dilution
by starlight in the host galaxy, the X--ray selection of AGN is not affected
by some of the biases of the optical selection \citep[see
e.g.][]{bra05}. The above mentioned optical QSO correlation functions have
been determined from surface densities of less than 30 AGN deg$^{-2}$. The
most recent deep optical surveys reach surface densities of $\sim$250
deg$^{-2}$ in the case of COMBO--17 (Classifying Objects by Medium-Band
Observations -- a spectrophotometric 17--filter survey) \citep{wol03}, and
$\sim$470 deg$^{-2}$ \citep{gav06} in the case of the Virmos--VLT Deep
Survey (VVDS).  The COMBO--17 survey contains 192 QSOs selected from
intermediate--band photometric redshifts with z$_{phot}>1.2$ and
$17<R_{vega}<24$ over 0.78 deg$^2$, while the VVDS contains 74
spectroscopically selected QSOs in a solid angle of 0.6 deg$^2$ with
$17.5<I_{AB}<24$.  The large factor between the X--ray and the optical
surface densities is crucial to map structures to a smaller scale than that
(1 $h^{-1}$Mpc) mapped with the current optical samples. Therefore, the most
sensitive spatial clustering analysis of AGN, especially at high redshift,
will be obtained by an X--ray survey, which is both wide and deep
simultaneously. In addition, an X--ray survey including also the hard (2--10
keV) band will give clustering information separately for the unabsorbed
(type--1) and the absorbed (type--2) AGN population and can thus directly
test whether the obscured and unobscured AGN originate from the same parent
population of galaxies, a crucial ingredient in AGN unification
scenarios. To study the 3D spatial correlation as a function of redshift and
source class, a few hundred spectroscopically identified AGN are required
per class and redshift shell in a contiguous sky area.  This is the main
science goal of the XMM--Newton COSMOS Survey.

There is very good evidence that AGN and galaxies co--evolve: the peaks of
AGN activity and star formation occur in the same redshift range (z =
1--2.5) and there is a similar dramatic decline towards low redshift. Black
hole mass and galactic bulge properties in the local universe are strongly
correlated \citep{kor01,mer01}. Therefore, the evolution of high--redshift
AGN can be utilized to study the formation and evolution of galaxies. The
observed redshift distribution of faint X--ray sources
\citep{cow03,has03,fio03,ued03} both in the soft and the hard bands, peaks
at z$\sim0.7$ and is dominated by relatively low luminosity objects (log
$L_X$ = 42--44 erg s$^{-1}$).  Recently, the evolution of the space density
of X--ray selected AGN has been derived \citep{has05} based on a sample of
$\sim$1000 type--1 AGN selected in the 0.5--2 keV band from ROSAT,
XMM--Newton and Chandra deep surveys with a very high completeness of
optical redshift identification ($>97\%$). A strong dependence of the AGN
space density evolution on X--ray luminosity has been found. The comoving
space density of high luminosity (log $L_X > 44$ erg s$^{-1}$) AGN rises
rapidly by more than a factor of 100 from z=0 to z$\sim$2, while the space
density of low luminosity AGN evolves significantly less rapidly in the same
redshift interval.  Moreover, there is a clear increase of the peak redshift
with increasing X--ray luminosity, both in the 0.5--2 keV and the 2--10 keV
band \citep{has03,fio03,ued03,has05}. Also, for the first time there is firm
evidence of a decline in the space density of lower--luminosity AGN towards
higher redshift.  This behaviour was not recognized in previous shallower
surveys, e.g. from ROSAT \citep{miy00}, and was not included in the XRB
synthesis models based on these \citep[e.g.][]{gil01}.

These new results paint a dramatically different evolutionary picture for
low--luminosity AGN compared to the high--luminosity QSOs. While the rare,
high--luminosity objects can form and feed very efficiently rather early in
the universe, with their space density declining more than two orders of
magnitude at redshifts below z=2, the bulk of the AGN has to wait much
longer to grow or to be activated, with a decline of space density by less
than a factor of 10 below a redshift of one. The late evolution of the
low--luminosity Seyfert population is very similar to that which is required
to fit the Mid--infrared source counts and background \citep{fra02} and also
the bulk of the star formation in the Universe \citep*{mad98}, while the
rapid evolution of powerful QSOs traces more closely the history of
formation of massive spheroids \citep{fra99}.  This kind of
anti--hierarchical Black Hole growth scenario is not predicted by most
semi--analytic galaxy evolution models based on Cold Dark Matter structure
formation \citep[e.g.][]{kau00,wyi03,crt05}. It could indicate two modes of
accretion and black hole growth with radically different accretion
efficiency \citep[see e.g. the models of][]{dim03,mer04,men04}.

The strong peak at z=0.7 has been somewhat contested
by \citet{tre04}, who noted that the decline at high redshift
is affected by a bias against the identification of distant
obscured AGN, while the obscured to unobscured AGN ratio at high-z
was still consistent with the value of $\sim$4:1 observed 
locally. Indeed $\sim$40\%  of the local Seyfert galaxies 
are Compton thick \citep{ris99}. They are missing from all current X--ray 
surveys, but should show up in mid-infrared surveys due to their
dust emission. On the other hand, the number of highly obscured 
objects missing in the current deep X--ray samples is limited
by the integrated constraints on the number counts, the hard X--ray and
soft gamma--ray background \citep{ued03} and the mass function of 
supermassive black holes in local galaxies \citep{mar04}. 
Including the observed decrease of obscuration with increasing X--ray 
luminosity \citep{ued03,has04}, there is an excellent consistency 
with all observed constraints \citep{ued03,tre05,gil06}.
 
While the evolution of AGN of all luminosity classes at $z<2$ has now been
firmly established from the rich X--ray samples discussed above, little is
known about the growth phase of high--mass black holes in the redshift range
3--6. Unlike in the optical and radio bands, where a clear decline of the
space density is observed in the redshift range 3--6 for the most luminous
QSOs \citep{fan01,wal05}, no decrease in the space density of luminous
X--ray QSOs is apparent up to z=4 \citep{has05}. Above that redshift, some
evidence of a decline has been seen by \citet{sil05} and \citet{has05}, but
the number of objects available so far is way too small to obtain accurate
quantitative information about the AGN growth phase, in comparison to the
optical QSO surveys. The total sample of AGN more luminous than logL$_X$=44
at z$>3$ collected from all ROSAT, Chandra and XMM--Newton surveys is just
$\sim30$ \citep{sil05,ste04,has05}, while in the COSMOS survey, once
completed, we expect of order 100 such objects, based on an extrapolation of
the most recent type--1 AGN X--ray luminosity function \citep{has05}.  If
large enough samples of AGN (several 100s per unit redshift) can be
spectroscopically identified and their optical/NIR morphology can be studied
with high angular resolution, then the well known correlations between
observable quantities of the host galaxies and nuclei, such as surface
brightness profile, effective radius and the stellar component velocity
dispersion, which in the local Universe is tightly correlated with the black
hole mass \citep{kor01,mer01}, open the possibility to extend these studies
to high redshift and truly determine the history of accretion in conjunction
with galaxy formation.

\begin{figure*}
\epsscale{1.0} 
\plotone{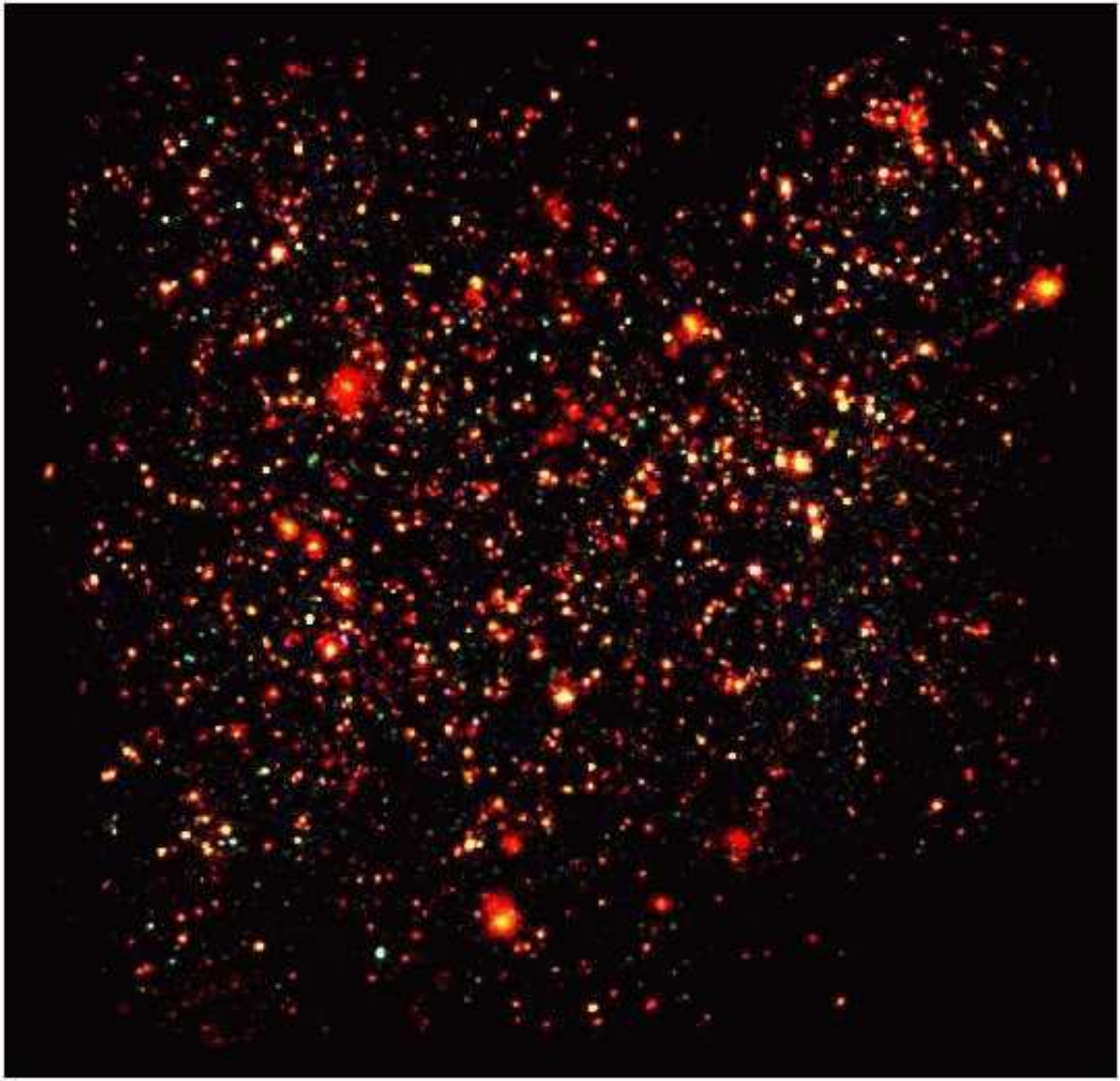}
\caption{False--color X-ray image of the XMM--Newton raster survey in the 
COSMOS field. For each of the 23 individual pointings with good time
intervals, raw images with 4\arcsec~ pixels have been accumulated in three
different energy bands:
0.5--2 keV (red), 2--4.5 keV (green) and 4.5--10 keV (blue). These images were
individually background--corrected and combined to a mosaic for all 23
pointings.  Then they have been filtered with a Gaussian kernel with a FWHM of 
4\arcsec.}  
\label{fig:img}
\end{figure*}

Clusters and groups of galaxies represent the second most abundant class of
objects identified in deep X-ray surveys. Due to their extended X-ray
emission they can be easily discriminated from point sources.  At redshifts
z$<$1 their rather low X--ray luminosities and temperatures correspond to
groups and low--mass clusters. A handful of clusters with redshifts z$>$1
has been discovered serendipitously in deep ROSAT \citep[e.g.][]{ros02} and
recently also XMM--Newton observations. The highest redshift is at
z$\sim$1.4 \citep{mul05}. At high luminosities the X--ray cluster population
shows mild cosmological evolution, consistent with the growth of structure
expected in a standard $\Lambda$CDM cosmology \citep{mul04a}. At fainter
X--ray fluxes considerable confusion can exist between the diffuse cluster
emission and faint AGN, either in the background or even as cluster members
\citep[e.g.][]{mar06}. Because of the strong cosmological AGN evolution, the
likelihood of AGN contamination substantially increases with redshift. High
angular resolution observations of some of the high redshift clusters and a
careful data analysis are thus important to understand possible biases
introduced by AGN contamination.

Another population of objects discovered in deep X-ray fields are stars
showing coronal emission, in particular G, K, and M-type stars with magnetic
flaring activity. While at bright fluxes (e.g. in the ROSAT All-Sky Survey)
coronal stars are quite abundant, the fraction of stars in the deep, high
Galactic latitude fields is only a few percent, mainly because the
observations are sensitive enough to reach beyond the typical stellar disk
populations.  Nevertheless, the statistics of stars in high latitude fields
can e.g.  constrain the decay of stellar magnetic activity on timescales of
several Gyr \citep{fei04}.

In this paper (paper I) we give an overview of the XMM--Newton survey in the
COSMOS field ({\bf XMM--Cosmos}), which is the basis for several
accompanying publications in the same journal issue, dealing with the {\em
''X--ray data and the logN--logS''} \citep[paper II,][]{cap06}, the {\em
''Optical identification and multiwavelength properties of a large sample of
X--ray selected sources''} \citep[paper III,][]{bru06}, the {\em ''X-ray
spectral properties of Active Galactic Nuclei''} \citep[paper IV,][]{mai06},
the {\em ''Angular Clustering of the X-ray Point Sources''} \citep[paper
V,][]{miy06}, and {\em ''Statistical properties of clusters of galaxies''}
\citep[paper VI,][]{fin06}.  Significant contributions from the XMM--Cosmos
survey are also contained in the papers on {\em ''A wide angle tail galaxy
in the COSMOS field: evidence for cluster formation''} \citep{smo06} and on
{\em ''A large-scale structure at z=0.73 and the relation of galaxy
morphologies to the local environment''} \citep{guz06}.

\begin{deluxetable*}{lccccccc}
\tablecaption{XMM--Newton Observing Log for the COSMOS Field\label{tab:obs}}
\tablehead{
\colhead{Target} & \colhead{\tablenotemark{a}} & \colhead{Obs. 
Date\tablenotemark{b}} &  \colhead{R.A.} & \colhead{Dec.} & \colhead{Exp.} & 
\colhead{GTI\tablenotemark{c}} & \colhead{BKG rate\tablenotemark{d} }\\
                 &  &                                    &\colhead{[hh mm ss]}  
&\colhead{[dd mm ss]}      & \colhead{[ks]}           &      \colhead{[ks]}                      
& \colhead{[cts/100s]} \\}
 \startdata
Field 1 & & 2004-12-11T13:45:22 & 10 02 25 & 02 44 15 & 31.0 & 25.8 &49.7  \\
Field 2 & & 2004-12-11T22:58:42 & 10 02 25 & 02 29 16 & 44.0 & 12.3 &48.7  \\
Field 3 & & 2005-05-14T03:40:20 & 10 02 28 & 02 10 55 & 32.1 & 26.9 &45.4  \\
Field 4 &*& 2004-11-21T05:34:24 & 10 02 25 & 01 59 15 & 31.0 & 23.2 &46.4  \\
Field 5 &*& 2004-11-21T14:47:44 & 10 02 25 & 01 44 15 & 31.0 & 23.4 &42.3  \\
Field 6 &*& 2004-05-30T01:11:41 & 10 01 25 & 02 40 56 & 32.2 & 19.7 &36.1  \\
Field 7 &*& 2003-12-06T01:58:02 & 10 01 25 & 02 29 16 & 34.5 & 28.9 &34.9  \\
Field 8 &*& 2004-11-17T22:11:57 & 10 01 25 & 02 14 16 & 53.0 & 24.5 &58.4  \\
Field 9 & & 2004-11-20T01:08:49 & 10 01 25 & 01 59 15 & 36.2 & 18.6 &44.4  \\
Field 10&*& 2004-11-22T00:01:05 & 10 01 25 & 01 44 15 & 45.5 & 11.5 &45.0  \\
Field 11& & 2004-12-01T23:46:01 & 10 00 25 & 02 44 15 & 44.1 & 17.5 &51.7  \\
Field 12&*& 2003-12-08T18:41:48 & 10 00 25 & 02 29 16 & 34.9 & 22.7 &33.3  \\
Field 13&*& 2003-12-10T11:46:18 & 10 00 25 & 02 14 16 & 31.8 & 22.9 &34.6  \\
Field 14&*& 2003-12-10T02:14:39 & 10 00 25 & 01 59 15 & 32.0 & 27.3 &33.8  \\
Field 15&*& 2004-11-19T15:55:29 & 10 00 25 & 01 44 15 & 30.1 & 18.3 &48.2  \\
Field 16& & 2004-12-06T01:17:33 & 09 59 25 & 02 42 36 & 40.1 &  0.0 &126.8 \\
Field 17& & 2004-12-11T04:15:21 & 09 59 25 & 02 29 16 & 31.2 & 26.8 &40.2  \\
Field 18&*& 2003-12-12T06:26:01 & 09 59 25 & 02 14 16 & 28.9 & 23.8 &35.9  \\
Field 19& & 2004-12-13T21:59:19 & 09 59 25 & 01 59 15 & 30.1 & 20.9 &36.7  \\
Field 20& & 2005-05-14T13:14:31 & 09 59 25 & 01 40 56 & 32.0 &  6.2 &15.6  \\
Field 21& & 2004-12-09T07:38:16 & 09 58 25 & 02 44 15 & 62.6 & 54.0 &39.4  \\
Field 22&*& 2004-11-03T06:24:56 & 09 58 25 & 02 29 16 & 31.0 & 25.3 &65.5  \\
Field 23& & 2005-05-09T19:23:50 & 09 58 25 & 02 10 57 & 31.0 &  8.1 &66.3  \\
Field 24& & 2005-05-10T04:37:10 & 09 58 25 & 01 55 58 & 31.0 & 15.5 &60.8 \\
Field 25& & 2005-05-10T14:20:30 & 09 58 25 & 01 40 56 & 32.0 &  0.0 &350.5 \\
\enddata
\tablenotetext{a}{Entries marked with an asterisk are part of the 12--field
sample, for which optical identifications are presented in \citet{bru06}.}
\tablenotetext{b}{Observation date yyyy-mm-ddThh:mm:ss}
\tablenotetext{c}{Net exposure time in PN--CCD good time intervals}
\tablenotetext{d}{Average value of quiescent particle background in the 
0.3--10 keV energy band after the cleaning}
\end{deluxetable*}

\section{XMM--Newton Observations}
 
We have surveyed an area of 1.4$\times$1.4 deg$^2$ within the region of sky
bounded by $9^h57.5^m<R.A.<10^h03.5^m$; $1^d27.5^m<DEC<2^d57.5^m$, which has
also been covered almost completely by the HST ACS \citep[1.67 deg$^2$,
see][]{lea06,sco06}.  The observations with XMM--Newton \citep{jan01} cover
a mosaic of 25 overlapping XMM pointings with a grid spacing of 15 arcmin,
i.e. about half of the size of the EPIC field of view.  The EPIC cameras
\citep{str01,tur01} were operated in the standard full--frame mode. The thin
filter was used for the PN camera and the medium filter for the MOS1 and
MOS2 cameras. Table \ref{tab:obs} gives a summary of the observations.

The PN and MOS data were preprocessed by the {\em XMM Survey Scientist
Consortium} \citep[SSC]{wat01} using the {\em XMM--Newton} Standard Analysis
System (SAS) routines. Because of software version changes, the data were
reprocessed homogeneously at MPE using the SAS version 6.5.0
\citep[see][]{cap06}.  A fraction of the observations was affected by high
and flaring background fluxes with count rates up to several hundred per
second, compared to a quiet count rate of several counts per second per
detector. The data were screened for low background intervals, rejecting
time intervals with more than $3\sigma$ enhancements in the background of
the individual observation \citep{cap06}.  The remaining good time intervals
for the PN--CCD detector added up to about 504 ksec (see
Tab. \ref{tab:obs}).  The requested exposure time for each pointing was 32
ksec, but some fields were observed longer due to scheduling
constraints. The actual exposure times for the three detectors, PN--CCD,
MOS1 and MOS2 are slightly different because of the different setup
times. Two pointings (Field 16 and 25) were completely lost due to high
background.

A number of hot pixels and hot columns were removed automatically from the
event lists by the standard analysis software. We searched for additional
hot pixels in images accumulated in detector coordinates, but did not find
any.  An exposure map for the whole mosaic of 23 pointings was calculated
for the combination of PN--CCD plus MOS1 and MOS2 cameras
(Fig. \ref{fig:exp}). The map shows residual structure due to the inter--CCD
gaps and dead pixels; however, the raster strategy with a 15 arcmin step
significantly reduces the residual structures.  We have been granted another
pass over the COSMOS Field with a total exposure time of 600 ksec in the
next XMM--Newton observing period (AO4), thus almost doubling the existing
dataset.  The remaining large scale inhomogeneities due to missing pointings
and variable net exposure times will be further smoothed out through the
detailed planning of the upcoming observations. Because the pointing grid in
AO4 will be shifted by 1 arcmin in both R.A. and Dec., we also expect an
additional smearing of the small--scale inhomogeneities. These AO4
observations have already started and have partly been used in the wavelet
analysis for diffuse sources in paper VI \citep{fin06}. The full set of
observations will be reported elsewhere after completion of the whole X--ray
survey.

\section{Image Analysis}

An Al--K$_\alpha$ line at 1.5 keV is present in both detector types. The PN
background spectrum shows in addition two strong copper lines at 7.4 and 8.0
keV, which are not present in the MOS background. PN photons in the energy
ranges 7.2--7.6 and 7.8--8.2 keV and all photons in the energy range
1.45--1.54 keV have therefore been neglected in the further analysis. For
each separate pointing, images in celestial coordinates with a pixel size of
4\arcsec~ have been accumulated in the 0.5--2, 2--4.5 and 4.5--10 keV bands
for all three detectors.  These are the energy bands, in which the
XMM--Newton Standard Analysis is performed and which have also been used in
the first XMM--Newton publication on the number counts in the Lockman Hole
\citep{has01}. These bands are also rather close to the classical 0.5--2,
2--10 and 5--10 keV bands used in the literature, so that, given the
energy--dependent sensitivity of imaging X--ray observatories, the counts to
flux conversion has rather small systematic errors \citep[see][]{has01}.
The images have been individually background--subtracted and then mosaiced
to the full raster pointing \citep[see][for more details]{cap06}.

Variations in the astrometric reference between different pointings are less
than 2\arcsec~ and have been corrected for the current analysis.  For each
pointing the brighter X--ray sources have been statistically correlated to
point--like optical objects (i.e. AGN or stars) using a maximum likelihood
algorithm ({\it eposcorr} in SAS) to check the astrometry of the individual
pointing images. Small corrections of order 2--3\arcsec~ in Right Ascension
and Declination were applied to the sky coordinates calculated from the
original WCS keywords before jointly analysing all pointings \citep{cap06}.
Combined PN+MOS1+MOS2 images were accumulated in the bands 0.2--0.5 keV
(very soft), 0.5--2 keV (soft), 2--4.5 keV (hard) and 4.5--10 keV (very
hard), respectively.  Fig. \ref{fig:img} shows the background--subtracted
image of all cameras combined in an X--ray false--color representation. The
red, green and blue colors refer to the soft, hard and very hard images,
respectively.

\section{Summary of results}

The source count rates from the 23--field COSMOS mosaic in the 0.5--2,
2--4.5 and 4.5--10 keV band have been converted to 0.5--2, 2--10 and 5--10
keV fluxes, respectively, and corresponding source counts have been derived
in \citet{cap06} together with the sky coverage for our observations.  In
this paper there is also an extensive discussion of systematic effects on
the source counts, like source confusion and the Eddington bias.  The
logN--logS relations are nicely consistent with previous results from ROSAT,
ASCA, BeppoSAX, Chandra and XMM--Newton. However, at intermediate fluxes
they have unprecedented statistical accuracy due to the large number of
objects involved. The total number of sources (point--like plus extended)
detected in the 0.5--2, 2--10 and 5--10 keV bands are 1307, 735 and 187,
respectively, and the corresponding flux limits are 0.7, 3.3 and 10 $\cdot
10^{-15}$ erg cm$^{-2}$ s$^{-1}$. The detection threshold has been set at a
likelihood value of $L=-ln(P_s)=6$, where $P_s$ is the probability for a
spurious source detection. This corresponds roughly to a Gaussian confidence
threshold of 4.5$\sigma$. The number of spurious sources in the total COSMOS
survey depends on the number of statistically independent trials, which can
only be calculated numerically for our detection scheme.  Detailed Monte
Carlo simulations of the XMM--Newton observations in the COSMOS field, using
the realistic exposure time and background levels \citep{cap06} show that at
this likelihood threshold about 33 spurious sources, i.e. about 2.4\%, are
expected among the whole X--ray catalog of the COSMOS field.  This is
consistent with similar simulations for the Lockman Hole \citep{brn06} and
is also confirmed by the statistics of optically empty error circles in
\citet{bru06}. The total number of different X--ray sources detected by the
standard likelihood analysis method is 1416. The number of sources detected
in the soft, hard and very hard band only are 676, 89 and 3,
respectively. The number of 1416 sources includes 26 objects, which are
found significantly extended by the maximum likelihood detection algorithm
\citep{cap06}.

First identifications with the optical catalogs from the CFHT Megacam,
Subaru SuprimCam and HST ACS catalogs as well as the UKIRT K--band catalog
are presented in accompanying paper III \citep{bru06}. Optical/NIR
candidates for the X--ray sources are identified using the {\em likelihood
ratio technique}.  This is a commonly used technique for associating objects
in two catalogues with each other \citep{sut92}. It compares the probability
that the object in question is the true counterpart, given the catalogued
position errors, with the probability of finding a chance object with this
magnitude at the position of this counterpart.  The large majority of the
X--ray sources can be readily identified with high likelihood using this
procedure. Only rather small fractions of the X--ray sources cannot be
uniquely identified, because the error circles either are empty down to the
sensitivity of the optical/NIR images (6\%) or contain multiple, about
equally likely counterparts (4\%).  Spectroscopic identifications of the
optical counterparts of the X--ray sources have started using the IMACS
spectrograph at the Magellan telescope \citep{imp06} and the VIMOS
spectrograph at the VLT in the zCOSMOS survey \citep{lil06}. Including about
40 spectroscopic identifications already in the literature, mainly from the
2dF and the SDSS surveys, a total of 377 X--ray sources could already be
spectroscopically identified in the 23 XMM--Newton fields discussed here
\citep[see also][]{bru06}.

\begin{figure}
{\epsscale{1.0} 
\centerline{\includegraphics[angle=-90,width=12cm]{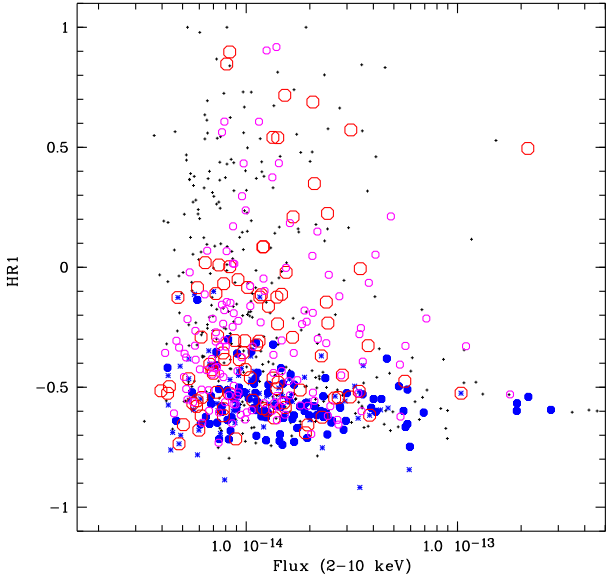}}
\caption{X--ray color--flux diagram of all sources detected 
in the 2--4.5 keV band in the COSMOS field. The 2--10 keV band flux (in erg
cm$^{-2}$ s$^{-1}$), calculated from the 2--4.5 keV band assuming a photon
index $\Gamma=1.7$, is plotted against the HR1 hardness ratio.  Different
symbols refer to the optical identification and classification of the
sources.  Filled circles are broad--line AGN spectroscopically identified in
the Magellan IMACS observations \citep{imp06} and from the literature.
Large open circles mark identified narrow emission line or absorption line
galaxies. Asterisks and small open circles correspond to optical
counterparts which are point--like or resolved, respectively, in the HST ACS
images of the first 12 optically identified fields
\citep[see][]{bru06}. Dots are X--ray sources in the remaining 11 fields,
for which optical identifications have not yet been attempted.}
\label{fig:FHHR}}
\end{figure}

Hardness ratios have been calculated from the counts in the three different
energy bands in the classical way: $$HR = (H-S)/(H+S)$$ HR1 refers to the
hard versus soft band and HR2 to the very hard versus the hard band.  Figure
\ref{fig:FHHR} shows an X--ray color--intensity diagram based on the 2--10
keV fluxes. Different symbols indicate different types of sources.  It is
easy to see that spectroscopically identified broad--line AGN (filled
circles), as well as X--ray sources with point--like ACS counterparts
(asterisks) only populate the soft part of the diagram, with hardness ratios
typically in the range $HR1<-0.2$. These are candidate type--1 AGN.  In this
figure there is a slight shift between the average locus of these unabsorbed
sources and the grid models, indicating that on average their spectra are
more complicated than simple power laws.  Sources spectroscopically
identified with narrow emission or absorption lines (large open circles), as
well as resolved ACS counterparts (small open circles) populate the whole
hardness ratio range. These are either candidate type--2 AGN (mainly those
with hard X--ray spectra) or low--luminosity type--1 AGN whose optical light
is dominated by the host galaxy.
 
\begin{figure}
\epsscale{1.0}
\centerline{\includegraphics[angle=-90,width=9cm]{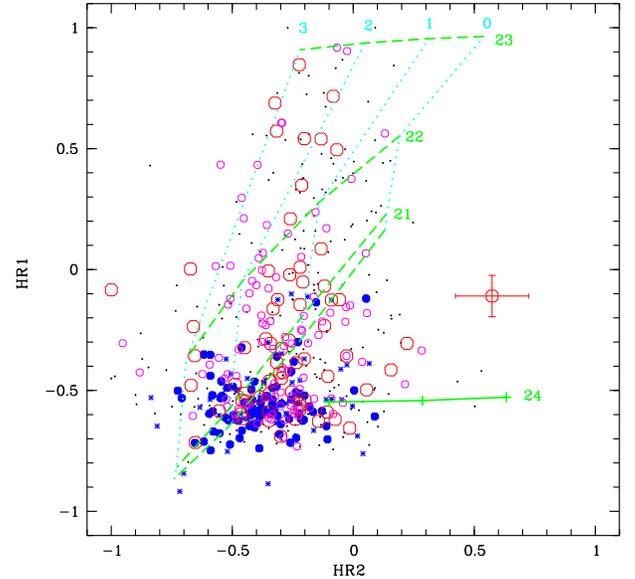}}
\caption{X--ray color--color diagram of the sources detected in the 
COSMOS field.  For clarity, only those sources are plotted, for which the
errors in HR1 and HR2 are less than 0.25. One representative set of error
bars has been plotted for the source with XID \#2608, which is discussed
specifically in the text.  The symbols are the same as in Figure
\ref{fig:FHHR}.  The grid lines refer to spectral models simulated using
XSPEC. Dashed and dotted lines show simple power law spectra with photon
indices $\Gamma$=0, 1, 2, 3 and intrinsic absorption (in the observed frame)
of log N$_H$= 21, 22, 23 cm$^{-2}$ indicated in the diagram. The grid line
for log N$_H$=0 cm$^{-2}$ is slightly below that for log N$_H$=21
cm$^{-2}$. The solid line corresponds to a reflection or leaky absorber
model with photon index of 1.7 with intrinsic absorption of N$_H$ =
10$^{24}$ cm$^{-2}$. The ticks along the line indicate the unabsorbed flux
percentage of 1, 3, and 10\% (from right to left). }
\label{fig:hr}
\end{figure}

\begin{figure*}
\epsscale{1.0}
\plotone{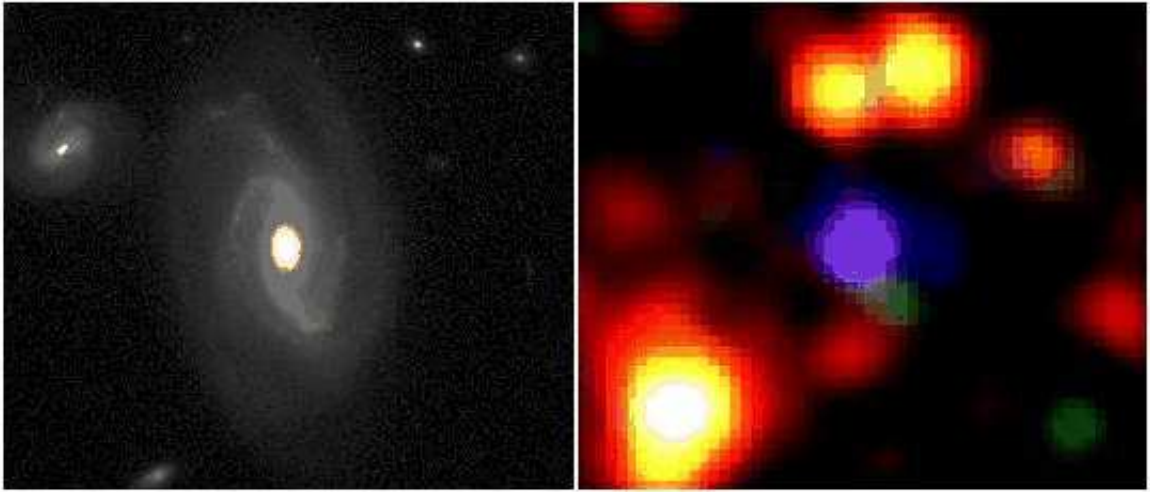}
\caption{Left: HST ACS cutout centered on the spiral galaxy 
2MASX J10014194+0203577 from the COSMOS ACS coverage \citep{sco06},
corresponding to XID \#2608, the hardest spectroscopically identified X--ray
source in the current XMM--Newton coverage of COSMOS. This galaxy has an ACS
I--band magnitude of 16.1 and a 2MASS K--band magnitude of 13.2. The
redshift measured by the SDSS is 0.1248. There is a nearby counterpart,
possibly weakly interacting. The image is 30\arcsec~ across.  Right: cutout
of the X--ray false color image in Fig. \ref{fig:img}, centered on XID
\#2608. The image is approximately 3\arcmin~ across. The peculiar spectrum
of the X--ray source is immediately apparent in the violet color, which
indicates contributions from the soft and the very hard band, but little
contribution from the hard band.}
\label{fig:acsxmm}
\vspace*{10mm}
\end{figure*}

Figure \ref{fig:hr} shows an X--ray color--color diagram for the same
sources. For clarity we plotted only sources with reasonably small errors in
the hardness ratio and which are significantly detected either in the hard
or the very hard band (see caption).  The grid of dashed and dotted lines,
as well as the single solid line show different power law models folded
through the instrument response of PN+MOS1+MOS2.  As already found in
\citet{has01} and \citet{del04}, type--1 AGN cluster in a well--defined
location in this diagram, corresponding to low intrinsic absorption and
spectral photon indices close to 1.7--2.  The fact that the median of the
type--1 AGN is shifted slightly to the right of the $N_H=0$ line may
indicate the presence of additional soft components in some of the spectra.
Type--2 AGN tend to have larger hardness ratios, especially in HR1,
corresponding to larger intrinsic absorption values. Interestingly, there is
a group of sources, putative type--2 AGN, which have quite hard colors in
the hard band (HR2), but hardness ratios consistent with unabsorbed sources
in the soft band, which do not fit any of the simple power law plus
absorption models. The solid model track shows, however, that their colors
can be reconciled with heavily absorbed low redshift sources with some
fraction (1--10\%) of unabsorbed flux leaking out. At higher redshifts, the
intrinsically absorbed continua move to a softer band, so that these objects
would shift to a different place in the color color diagram \citep{gua05}.
A single prototypical object (\#901) has already been discovered in the
XMM--Newton observations of the Lockman Hole \citep{mai02}. \citet{gua05}
have recently shown that bright, local Compton--thick Seyfert--2 galaxies
populate the same area in the color--color diagram. The large solid angle
coverage of the COSMOS field now clearly identifies them as a population of
the probably most absorbed objects in the field.  Using the tentative
criterion of HR2$>0.1$ and HR1$<-0.1$, a total number of 18 candidates can
be selected, which is $\sim3\%$ of the total number of objects shown in
Fig. \ref{fig:hr}, a fraction consistent with that detected in the Lockman
Hole \citep{has01}.  \citet{mai06} elaborate on the spectral analysis of
individual sources, concentrating on the brighter spectroscopically
identified sources in the COSMOS field.

A particularly interesting source is XID \# 2608, the source with the
largest HR2 ratio among those spectroscopically identified in Figure
\ref{fig:hr}.  It is a galaxy with a redshift z=0.125 from the SDSS archive
and an I--band magnitude of 16.11 \citep{bru06}. Combining all three EPIC
cameras, it has 72.4, 43.1 and 133.3 net counts in the soft, hard and very
hard band, respectively, and thus a very peculiar X--ray spectrum. The 5--10
keV flux derived from the counts in the very hard band is $\sim 4 \times
10^{-14}$ erg cm$^{-2}$ s$^{-1}$, corresponding to a 5--10 keV luminosity of
$1.3 \times 10^{43}$ erg s$^{-1}$.  Fig. \ref{fig:acsxmm} shows a cutout
from the ACS mosaic on the left and a zoom into the X--ray false color image
on the right. The peculiar X--ray spectrum of this possibly weakly
interacting spiral galaxy is immediately apparent from its strange violet
X--ray color. A high quality optical spectrum for this source is available
in the SDSS archive and has been analysed by \citet{kau03} as part of their
massive spectral analysis of all SDSS emission line galaxies. It has been
classified by these authors as a narrow--line Seyfert galaxy with moderate
power (logL[OIII]=6.45). The host galaxy is rather massive with a stellar
mass around $(2-3)~10^{11}$ M$_\odot$ and a stellar velocity dispersion of
$\sigma_v\sim$160 km/s.  Despite its absence in the total spectrum, a
significant H$\beta$ emission line is found by these authors in the residual
AGN spectrum after subtraction of the appropriate galaxy spectrum. The
diagnostic line ratios log [OIII]/H$\beta$=0.61 and log [NII]/H$\alpha$=0.12
put this object among the bulk of the normal Seyfert--2 galaxies.  With its
X--ray and [OIII] luminosity and its peculiar X--ray spectrum, the object
can be identified as a heavily absorbed, moderately powerful Seyfert--2
galaxy, with a soft component from a leaky absorber, a scatterer/reflection
nebula \citep{elv83} or a circumnuclear starburst region.

First science results from the XMM--Newton survey of the COSMOS field have
also been derived in the realm of large scale structure and clusters of
galaxies. A survey of clusters and groups of galaxies, based on extended
X--ray sources detected by a wavelet algorithm in the dataset of 36
XMM--Newton pointings and overdensities in optical photometric redshift
catalogs has detected 72 candidate groups and clusters over the whole field
\citep{sco06,fin06}, of which the 19 most massive ones have independently
been confirmed through a weak lensing analysis \citep{tay06}.  These results
provide first constraints on the number counts and mass function of groups
and clusters in this so far unexplored flux range.  A particularly
interesting group of clusters has been found at a redshift around z=0.73
\citep{guz06}. A first angular correlation function of the active galactic
nuclei in the 23--field survey finds a significant signal in the range
$\sim$0.5--20\arcmin~ in all three energy bands 0.5--2, 2--4.5 and 4.5--10
keV \citep{miy06}.

\section{Public Data Release / Acknowledgements}

The COSMOS XMM--Newton data are publicly available in staged releases
through the web sites at MPE:
\url{http://www.mpe.mpg.de/XMMCosmos/23fields/}, and the one at at
IPAC/IRSA: \url{http://irsa.ipac.caltech.edu/data/COSMOS/}.  The XMM--Newton
pipeline processed data are of course also available in the XMM--Newton
archive.

\acknowledgments

This work is based on observations obtained with XMM--Newton, an ESA science
mission with instruments and contributions directly funded by ESA Member
States and the US (NASA). In Germany, the XMM--Newton project is supported
by the Bundesministerium f\"ur Bildung und Forschung/Deutsches Zentrum f\"ur
Luft und Raumfahrt, the Max--Planck Society, and the
Heidenhain--Stiftung. Part of this work was supported by the Deutsches
Zentrum f\"ur Luft-- und Raumfahrt, DLR project numbers 50 OR 0207 and 50 OR
0405. We gratefully acknowledge the contributions of the entire COSMOS
collaboration consisting of more than 100 scientists. More information on
the COSMOS survey is available at
\url{http://www.astro.caltech.edu/~cosmos}. This research has made use of
the NASA/IPAC Extragalactic Database (NED) and the SDSS spectral archive. We
thank Guinevere Kauffmann and Jarle Brinchmann for help with the object
\#2608. We acknowledge helpful comments from an anonymous referee.

\end{document}